\documentclass[twocolumn,showpacs,aps,prl]{revtex4-1}
\usepackage{graphicx}
\usepackage{color}

\newcommand{\sub}[1]{_{\mbox{\scriptsize{#1}}}}  %non-italicized subscript

\begin{document}

\title{Longitudinal spin diffusion in a nondegenerate trapped $^{87}$Rb gas}
\author{D.~Niroomand, S.D.~Graham, and J.M.~McGuirk}
\affiliation{Department of Physics\\
Simon Fraser University, Burnaby, British Columbia V5A 1S6, Canada}
\date{\today}

\begin{abstract}
Longitudinal spin diffusion of two pseudo-spin domains is studied in a trapped $^{87}$Rb sample above quantum degeneracy, and the effect of coherence in the domain wall on the dynamics of the system is investigated. Coherence in a domain wall leads to transverse-spin-mediated longitudinal spin diffusion that is slower than classical predictions, as well as altering the domains' oscillation frequency. The system also shows an instability in the longitudinal spin dynamics as longitudinal and transverse spin components couple, and a conversion of longitudinal spin to transverse spin is observed, resulting in an increase in the total amount of coherence in the system.
\end{abstract}

%51.10.+y	Kinetic and transport theory of gases
%51.20.+d   diffusion in gases
%67.85.-d	Ultracold gases, trapped gases
%75.40.Gb   spin diffusion

\maketitle

The development of ultracold trapped atomic systems of fermions and bosons has opened up new paths of study for transport phenomena. In particular, theoretical and experimental studies of spin transport have been of recent interest in a number of systems including strongly interacting degenerate Fermi gases near unitary \cite{kohl2013, bardon2014transverse, trotzky2014observation}, dilute noncondensed bosonic gases close to quantum degeneracy \cite{mcguirk2002sw, mcguirk2011localized}, and ultracold gases in optical lattices \cite{hild2014far, heinze2013engineering}.

Quantum degeneracy effects can dramatically modify spin transport in spin-polarized systems, for example leading to phenomena such as the Leggett-Rice effect \cite{Leggett1968}. Though spin is a fundamentally quantum property, classical Boltzmann theory was expected to describe spin transport in dilute nondegenerate gases accurately, since quantum degeneracy effects are typically insignificant. Surprisingly in 1982 Bashkin \cite{bashkin1981spin} and Lhuillier and Lalo{\"e} \cite{lhuillier1982transport} independently predicted that transport in quantum gases can lead to macroscopic collective behavior even in nondegenerate systems with spin-independent interactions. The driving process for this collective behavior is the identical spin rotation effect (ISRE), in which symmetrization of the two-body wave function in binary collisions between identical particles leads to coherent exchange interactions when the thermal de Broglie wavelength is larger than the two-body interaction length \cite{lhuillier1982transport}. Examples of such collective behavior include spin waves \cite{mcguirk2002sw}, spin self-rephasing \cite{reichel2010coherence}, and non-conservation of transverse spin \cite{ragan1997castaing}.

Diffusion is also strongly modified by quantum effects, leading to nonlinear and anisotropic spin diffusion \cite{lhuillier1982transport2,mullin2006spin} and universal spin dynamics \cite{kohl2013}. Near unitarity, experiments on two spin domains in a strongly interacting Fermi gas showed reversal of spin currents \cite{sommer2011universal}, resulting in quantum-limited spin diffusion. Also in unitary Fermi gases, further quantum modifications to diffusive timescales were observed \cite{kohl2013,bardon2014transverse}. In \cite{hild2014far}, patterned spin textures were used to study diffusion in various geometries, showing diffusive behavior in one dimension (1D), but large departures from classical transport in 2D.

In the nondegenerate case of a 1D two-domain spin structure studied here, quantum modifications to diffusion appear as a decrease in the oscillation rate of spin domains in a harmonic potential and an increase in the longitudinal spin diffusion time, through coupling to the transverse spin. This effect shows sensitivity to the degree of coherence in the domain wall between spin domains, which highlights the quantum mechanical nature of the dynamics more clearly. In this Letter we demonstrate the alteration of longitudinal spin diffusion timescales depending on the initial amount of the coherence in the domain wall and also report observation of a longitudinal spin instability, which leads to an increase in the total amount of coherence in the system.

The mechanism for coherent atom-atom scattering is the ISRE, where exchange scattering between indistinguishable particles leads to precession of each atom's spin about their combined spin. The theoretical description of spin dynamics in nondegenerate quantum systems was developed using semiclassical kinetic theory \cite{lhuillier1982transport2, fuchs2002internal,nikuni2002linear, oktel2002internal, williams2002longitudinal}. We adopt the notation of \cite{nikuni2002linear} and describe the time evolution of the spin density distribution, $\vec{\sigma}(p,z,t)$, with a 1D quantum Boltzmann equation that includes the ISRE via a spin-torque term:
\begin{equation}
    \partial_t \vec{\sigma}(p,z,t)+\partial_0\vec{\sigma}(p,z,t)-\vec{\Omega}\times\vec{\sigma}(p,z,t)=\frac{\partial\vec{\sigma}}{\partial t}|_{1D},
    \label{eq:Boltzman}
\end{equation}
where $\partial_0=\frac{p}{m}\frac{\partial}{\partial z}-\frac{\partial U\sub{ext}}{\partial z}\frac{\partial}{\partial p}$ and $\vec{\Omega} = (U\sub{diff}\,\hat{u}_\parallel + g\vec{S})/\hbar$. The mean-field interaction strength for two scattering particles with $s$-wave scattering length $a$ and mass $m$ is $g=\frac{4\pi {\hbar}^2 a}{m}$, and $\hat{u}_\parallel$ is the longitudinal unit vector on the Bloch sphere. The experimentally observable quantity is the spin $\vec{S}(z,t)=\int dp\,\vec{\sigma}(p,z,t)/2\pi\hbar$. $\vec{S}$ contains the longitudinal spin $S_{\parallel}$ and the transverse spin $\vec{S}_\perp$, with magnitude $S_{\perp}$ (i.e.~the spin coherence). $U\sub{ext}$ and $U\sub{diff}$ are the trapping potential and differential potential experienced between the two states (due to mean-field and Zeeman shifts). Lastly, the right-hand side of Eq.~\ref{eq:Boltzman} represents damping due to elastic collisions \cite{nikuni2002linear}.

We create a two-domain structure of ``up-down'' pseduo-spin states in a nondegenerate gas of $^{87}$Rb atoms trapped in an external harmonic potential and observe the relaxation to equilibrium. The trap is axisymmetric, with trapping frequencies of $2\pi\times(6.7,260,260)$~Hz. Due to rapid averaging from the high radial frequency, dynamics can be treated effectively as one-dimensional. The pseudo-spin doublet consists of two magnetically trapped hyperfine ground states of $^{87}$Rb ($|1\rangle\equiv|1,-1\rangle$ and $|2\rangle\equiv|2,1\rangle$), coupled via a two-photon microwave transition. (See \citep{mcguirk2010optical,niroomand2013observation} for details.)

\begin{figure}[ht]
    \includegraphics[width=\linewidth,clip]{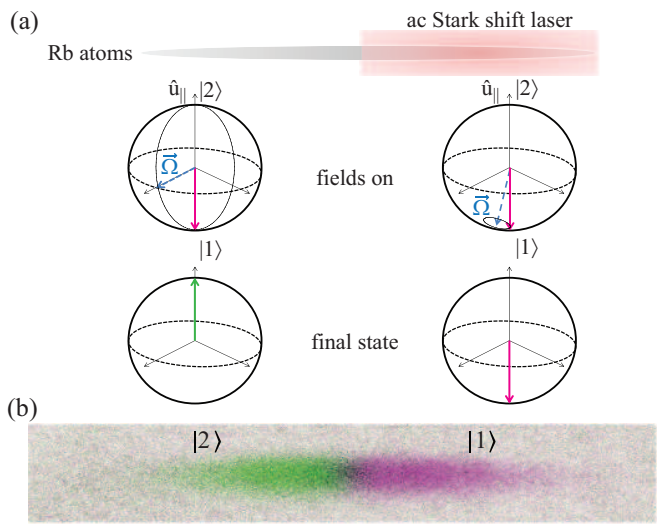}
    \caption{(a) Schematic of the domain preparation, showing the patterning laser and accompanying spin evolution in the Bloch sphere representation (longitudinal axis $\hat{u}_\parallel$ and transverse equatorial plane). Applying a $\pi$-pulse to atoms initially in $|1\rangle$ while illuminating them with the masked laser transfers atoms in the unilluminated half (left) of the distribution to $|2\rangle$ while leaving atoms in the illuminated half unchanged (right). (b) Composite image of atoms in $|1\rangle$ (purple) and $|2\rangle$ (green) following initialization.}
    \label{spinprofile}
\end{figure}

We prepare spin domains with optical patterning and microwave pulse techniques (see Fig.~\ref{spinprofile}) \citep{mcguirk2010optical,niroomand2013observation}. Magnetically trapped $^{87}$Rb is evaporatively cooled to just above quantum degeneracy (temperature $T\sim 650$~nK and peak density $n \sim 2.6\times10^{13}$~cm$^{-3}$). Initially in state $|1\rangle$, a microwave $\pi$-pulse transfers atoms to $|2\rangle$, except where illuminated with an off-resonant, partially masked laser beam that creates an optical step potential on top of the trapping potential. The masked laser ac Stark shifts atomic energy levels out of resonance by $\sim50\Omega\sub{R}$. This procedure creates domains of spin up and down separated by a helical domain wall in which the spin vector is coherently twisted, while remaining fully polarized.

The longitudinal spin projection is measured by directly measuring the population in states $|1\rangle$ and $|2\rangle$ using absorption imaging. Each image is divided into equally spaced axial bins, and the local population difference is given by $S_\parallel (z)=\frac{N_2(z)}{N_2^{\rm{tot}}} -\frac{N_1(z)}{N_1^{\rm{tot}}}$ with $N_i^{\rm{tot}}=\sum\limits_{z} N_i(z),\,(i=1,2)$ summed over all axial bins. Ramsey-type experiments measure the transverse spin component. We normalize transverse spin by the magnitude of a fully coherent sample with no spin inhomogeneities, $c(z,t) \equiv S_\perp(z,t)/S_\perp^{\mbox{\scriptsize{max}}}$, and the total coherence in the sample is $c\sub{tot}(t) \equiv \int c(z,t)\,dz$.

\begin{figure}[ht]
\centering
    \includegraphics[width=\linewidth,clip]{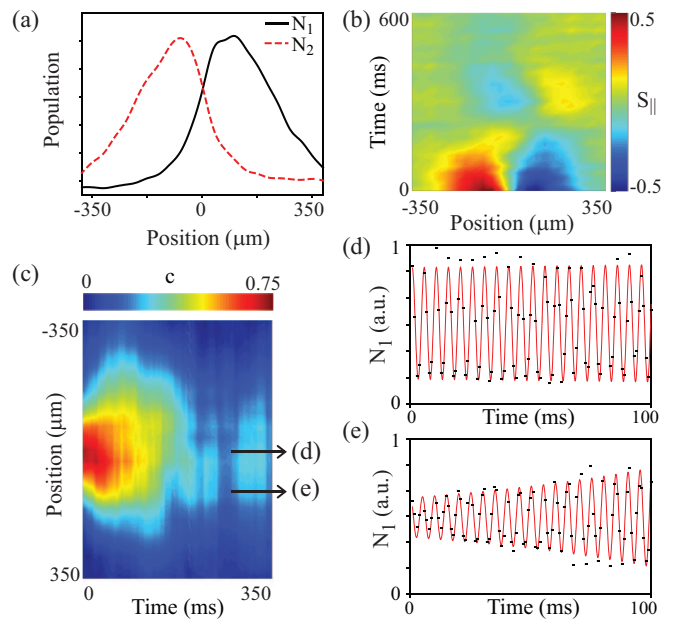}
    \caption{Time evolution of spin domains with $c\sub{init}=0.74$. (a) Typical initial spin distribution of $|1\rangle$ (solid line) and $|2\rangle$ (dashed line), used to infer $S_\parallel$. (b) Evolution of $S_\parallel$. (c) Evolution of transverse spin amplitude, $c(z,t)$. Coherence is initially confined to the domain wall, in a region less than one-third the axial full-width half-maximum of the atomic distribution, $\sim0.3\,\mbox{w}\sub{ax}$, but rapidly spreads toward the edges of the cloud, covering a region of the size $\sim0.7\,\mbox{w}\sub{ax}$ by 200~ms. Ramsey fringes at (d) the trap center $z=0$ and (e) $z=0.3\,\mbox{w}\sub{ax}$ as shown by arrows in (c), with a sinusoidal fit to guide the eye.}
    \label{rawdata}
\end{figure}

Figure~\ref{rawdata} shows the typical time evolution of the two-domain spin structure. The longitudinal spin component $S_\parallel$ exhibits spin domain oscillations as well as diffusion of the longitudinal spin gradient, which damps the oscillations. These measurements reveal that longitudinal domains persist for much longer than expected for classical longitudinal diffusion times ($\sim25$~ms \cite{nikuni2002linear}) and oscillate much slower than trap oscillations (Fig.~\ref{rawdata}(b)). A coherent domain wall has a stabilizing effect, which agrees with theoretical predictions and previous experiments in spin-polarized gases \cite{nunes1992spin, akimoto1991nonlinear, konig1995spin, ragan1997castaing}. We study this effect in more detail by controllably tuning the amount of initial coherence in the domain wall.

The more striking behavior occurs in the time evolution of the transverse spin component. Fig.~\ref{rawdata}(c) shows the spatiotemporal evolution of the amplitude of $S_\perp$ (coherence) extracted from Ramsey measurements (Fig.~\ref{rawdata}(d-e)). Ramsey fringe contrast rapidly rises within longitudinal spin domains, where the initial fringe amplitude is small (Fig.~\ref{rawdata}(e)), showing the spread of coherence toward the cloud edges.

\begin{figure}[t]
    \includegraphics[width=\linewidth,clip]{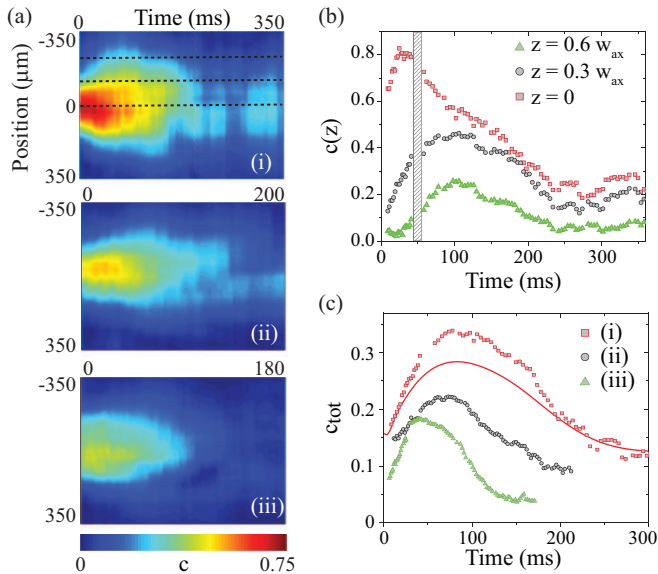}
    \caption{(a) Time evolution of $S_\perp$ for different domain wall coherences, from top to bottom $c\sub{init} = 0.71,\ 0.51,\ 0.30$ respectively in the cloud center. (b) Time evolution of $S_\perp$ in different spatial regions, denoted by $z = (0,\ 0.3,\ 0.6)\times \mbox{w}\sub{ax}$ [dotted lines in (i)]. The dashed box indicates a change in sampling rate to measure both the fast initial rise of coherence and longer relaxation to equilibrium. (c) Time evolution of total ensemble coherence for $c\sub{init}$ shown in (a). The solid curve is a numerical simulation of Eq.~\ref{eq:Boltzman} for (i), showing good qualitative agreement. Quantitative agreement is limited by challenges extracting Ramsey fringe amplitudes in the presence of noise, dephasing, and small signal in cloud wings.}
    \label{tr}
\end{figure}

Spreading of coherence could result from $S_\perp$ diffusion, but the total transverse spin magnitude in the gas increases (Fig.~\ref{tr}(d)). This increase implies the appearance of coherence near the cloud edges is not due to spin diffusion, but rather to conversion of $S_\parallel$ into $S_\perp$ as a result of an instability in the longitudinal spin current. This effect has been observed in spin-polarized gas systems \cite{nunes1992spin,akimoto1991nonlinear} and was described as an experimental manifestation of the Castaing instability \cite{ragan1997castaing}. Although the trapped atomic system possesses different experimental parameters, the physics governing the phenomenon is similar \cite{ragan2004leggett, kuklov2002precessing, fuchs2003castaing}. If transverse spin is confined in the domain wall, it dephases and the $S_{\parallel}$ gradient decays via ordinary longitudinal diffusion, since the kinetic equation decouples for longitudinal and transverse spin components \cite{ragan2004leggett, meyerovich1989quantum}. However, if the ISRE is large enough (i.e.~large Leggett-Rice parameter), the longitudinal spin current becomes unstable, coupling between longitudinal and transverse spin dynamics occurs, and the $S_\parallel$ gradient decays via transverse diffusion across the coherent domain wall. Thus transverse-spin-mediation of longitudinal diffusion increases the timescale for longitudinal diffusion to be comparable with slower transverse diffusion times, which are slower due to coherent spin interactions absent from longitudinal spin dynamics.

To study the effect of initial coherence in the domain wall, we initialize domain walls with different $S_\perp$ and observe their relaxation. Coherence is controlled via the same off-resonant laser that prepares the spin domains. A short light pulse creates a nonuniform differential atomic potential, whose inhomogeneity induces rapid dephasing of $\vec{S}_\perp$ in the domain wall. The degree of coherence is controlled by the pulse length ($\leq0.6$~ms) and measured via Ramsey spectroscopy. This forced dephasing procedure can reduce the initial coherence by over 70\%.

\begin{figure}[t]
    \includegraphics[width=\linewidth,clip]{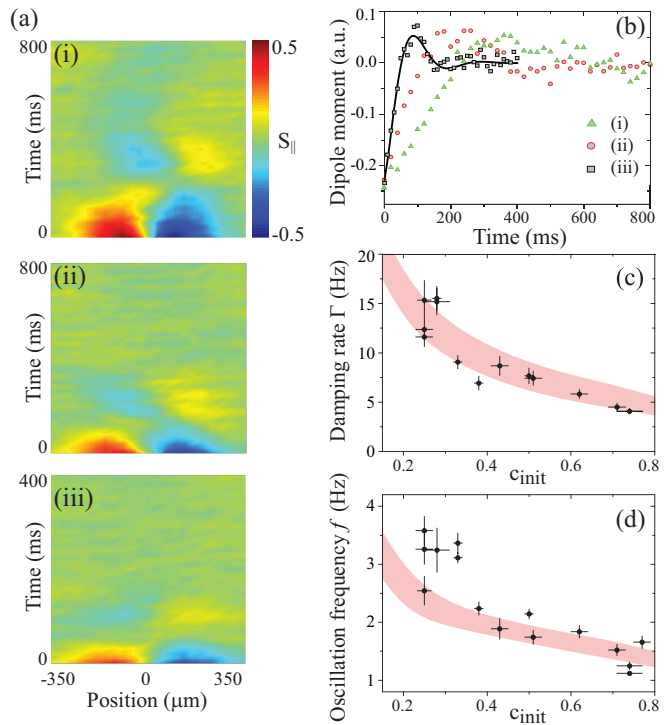}
    \caption{(a) Time evolution of $S_\parallel$ for different initial domain wall coherences, from top to bottom $c\sub{init} = S_\perp/S_\perp^{\mbox{\scriptsize{max}}} = 0.74,\ 0.51,\ 0.28$ respectively in the cloud center. (b) Dipole moment time evolution calculated from (a), with a representative decaying sinusoidal fit to (iii) $c\sub{init} =0.28$. (c) Damping rate and (d) oscillation frequency of the dipole moment for different $c\sub{init}$. Error bars correspond to fit uncertainties for dipole moment oscillations and $c\sub{init}$ measurements. The shaded band is the result of numerical simulations of Eqn.~\ref{eq:Boltzman}.}
    \label{lg}
\end{figure}

Figure \ref{lg}(a) shows the longitudinal time evolution for different initial degrees of coherence. Both the frequency of longitudinal spin domain oscillations as well as longitudinal spin diffusion rates decrease as domain wall coherence is increased. Dynamics for a two-domain structure are dominated by the dipole mode, which we isolate via the dipole moment of the spin distribution, $\left\langle z S_\parallel(z,t)\right\rangle _n$, where $\left\langle ... \right\rangle _n$ denotes a density-weighted average to compensate for low signal-to-noise ratio at the distribution edges. Figure~\ref{lg}(b) shows evolution of spin-dipole moments for the coherences in Fig.~\ref{lg}(a). The frequency $f$ and damping rate $\Gamma$ of longitudinal oscillations are extracted from these moments.

We repeat the experiment for different initial amounts of coherence, $c\sub{init}$; calculate the time-dependent dipole moment; and extract $f$ and $\Gamma$ from decaying sinusoidal fits to these oscillations. Figure~\ref{lg}(c) and (d) summarize the results. Both $\Gamma$ and $f$ decrease as $c\sub{init}$ increases, showing the stabilizing effect of a coherent domain wall. The oscillation frequency is primarily controlled by two competing factors: trapping potential oscillations and mean-field-induced spin rotation. In the incoherent limit, $f$ approaches the trapping frequency, as the system approaches a mixture of distinguishable classical ideal gases that diffuse according to classical Boltzmann theory. The large $c\sub{init}$ limit entrains longitudinal diffusion with slower transverse diffusion.

We compare these measurements with transport theory by numerically simulating Eqn.~\ref{eq:Boltzman} (see \cite{mcguirk2010optical} for details). The shaded regions in Fig.~\ref{lg}(c) and (d) represent one-sigma confidence bands from Monte Carlo simulations of Eqn.~\ref{eq:Boltzman}, including statistical fluctuations in $n$, $T$, and domain wall size, as well as a systematic density calibration uncertainty. The data agrees well with theoretical predictions. Discrepancies in $f$ at low coherence is likely due to challenges in fitting critically damped oscillations where the quality factor drops. Overdamping should occur for $c\sub{init}<0.2$, but reducing coherence to this level without altering the longitudinal spin domains is challenging.

The effect of the spin instability can be seen more clearly in the transverse spin dynamics (Fig.~\ref{tr}). The time evolution of $S_\perp(z,t)$ for different initial domain wall coherence is shown in Fig.~\ref{tr}(a). The rise and spread of coherence for different $c\sub{init}$ shows similar spatial behavior, but $S_\perp$ persists longer when there is more coherence in the domain wall initially. Figure~\ref{tr}(b) shows the evolution of coherence in different regions of the cloud for the high $c\sub{init}$ preparation (Fig. \ref{tr}a(i)).

Figure \ref{tr}(c) shows total ensemble coherence, $c\sub{tot}(t)$, for different $c\sub{init}$, calculated by summing the coherence across the cloud. Coherence rapidly rises followed by a gradual decrease, but with different timescales for different $c\sub{init}$. Maximum total coherence across the cloud depends on the initial coherence, and the time to reach maximum coherence also increases as $c\sub{init}$ increases. These results highlight the transverse source of enhanced lifetimes of longitudinal spin domains with coherent domain walls. The presence of coherence in domain walls links diffusion timescales and increases longitudinal diffusion times to be as long as transverse diffusion times.

%%%%%%%%%%%%%%%%%%%%%%%%%%%%%
%%%%%%%%%%%%%%%%%%%%%%%%%%%%%
Our results validate the quantum Boltzmann equation used to simulate the domain wall spin dynamics. As the region of experimental interest lies between the collisionless and hydrodynamic regimes (Knudsen number $\sim 1.2$), no analytic approximation is readily available for the instability criterion, such as was done in \cite{kuklov2002precessing}. However, there are two primary requirements for driving the instability. First and most important is that the ratio of exchange scattering to elastic scattering rates must be approximately greater than 1 (quantified by the Leggett-Rice parameter $\mu$ \cite{fuchs2003castaing}, $\sim 3$ for this experiment), lest collisions damp spin currents before they can grow large. The second requirement is that the number of exchange collisions while traversing the domain wall be small; if not, a spin crossing the gradient will smoothly rotate from one domain to the other, similar to adiabatic following.

\begin{figure}[t]
    \includegraphics[width=\linewidth,clip]{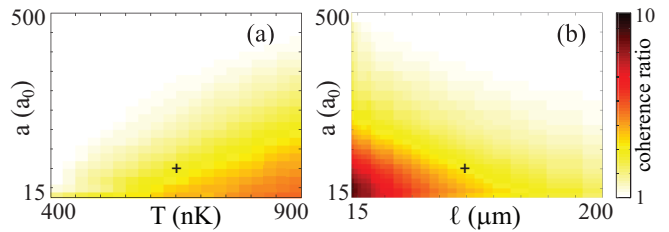}
    \caption{Instability onset seen in the ratio of maximum ensemble coherence to initial coherence, $c\sub{tot}^{\rm max}/c\sub{tot}^{\rm init}$, from simulations of Eq.~\ref{eq:Boltzman}, plotted logarithmically to highlight instability threshold. We vary $a$ and (a) $T$ or (b) $\ell$ with other parameters set to conditions of Fig.~\ref{lg}(a)(ii). Since $\mu\propto 1/a$, increasing $a$ diminishes the instability, as collisional damping grows faster than exchange scattering. Decreasing $T$ or increasing $\ell$ stabilizes the spin current as expected, through increasing adiabatic rotation during domain wall crossing. Crosses represent experimental conditions used here.}
    \label{sims}
\end{figure}

We extend these simulations to a broader parameter space than can be achieved in our system to explore further the crossover from classical to quantum diffusion, by numerically varying $T$, $n$, $a$, and domain wall size, $\ell$. Figure~\ref{sims} shows the effects of $T$, $a$, and $\ell$ on the spin instability. Because of near-critical damping and spatially varying density (thus collision rates), the system exhibits a smooth transition between diffusive regimes. Furthermore, since $\mu$ is independent of $n$, density has little effect on the instability within the range between collisionless and hydrodynamic behavior, and confinement plays little role so long as $\ell \ll \mbox{w}\sub{ax}$.

%%%%%%%%%%%%%%%%%%%%%%%%%%%%%
%%%%%%%%%%%%%%%%%%%%%%%%%%%%%

In conclusion, our results reveal the stabilizing effect of coherence in a domain wall, leading to an increase in longitudinal domain lifetimes. These effects are explained by coupling between longitudinal and transverse spin dynamics, which induces transverse-channel-mediated longitudinal spin diffusion. While not a direct measure of anisotropic diffusion, alteration of diffusive timescales by adjusting only the longitudinal-transverse spin coupling (via the coherence) is strongly suggestive of anisotropic diffusion. An accompanying rise in the total coherence indicates a conversion of longitudinal spin into transverse spin as a result of an instability in the longitudinal spin dynamics. Longer longitudinal domain lifetimes result from more coherence in the domain walls, and suppression of this instability leads to even longer longitudinal diffusion times. Our results agree well with numerical simulations and emphasize the importance of quantum effects in spin transport in trapped ultracold bosonic systems, even above quantum degeneracy.

\bibliography{spindiffusion}

%merlin.mbs apsrev4-1.bst 2010-07-25 4.21a (PWD, AO, DPC) hacked
%Control: key (0)
%Control: author (8) initials jnrlst
%Control: editor formatted (1) identically to author
%Control: production of article title (-1) disabled
%Control: page (0) single
%Control: year (1) truncated
%Control: production of eprint (0) enabled
\begin{thebibliography}{28}%
\makeatletter
\providecommand \@ifxundefined [1]{%
 \@ifx{#1\undefined}
}%
\providecommand \@ifnum [1]{%
 \ifnum #1\expandafter \@firstoftwo
 \else \expandafter \@secondoftwo
 \fi
}%
\providecommand \@ifx [1]{%
 \ifx #1\expandafter \@firstoftwo
 \else \expandafter \@secondoftwo
 \fi
}%
\providecommand \natexlab [1]{#1}%
\providecommand \enquote  [1]{``#1''}%
\providecommand \bibnamefont  [1]{#1}%
\providecommand \bibfnamefont [1]{#1}%
\providecommand \citenamefont [1]{#1}%
\providecommand \href@noop [0]{\@secondoftwo}%
\providecommand \href [0]{\begingroup \@sanitize@url \@href}%
\providecommand \@href[1]{\@@startlink{#1}\@@href}%
\providecommand \@@href[1]{\endgroup#1\@@endlink}%
\providecommand \@sanitize@url [0]{\catcode `\\12\catcode `\$12\catcode
  `\&12\catcode `\#12\catcode `\^12\catcode `\_12\catcode `\%12\relax}%
\providecommand \@@startlink[1]{}%
\providecommand \@@endlink[0]{}%
\providecommand \url  [0]{\begingroup\@sanitize@url \@url }%
\providecommand \@url [1]{\endgroup\@href {#1}{\urlprefix }}%
\providecommand \urlprefix  [0]{URL }%
\providecommand \Eprint [0]{\href }%
\providecommand \doibase [0]{http://dx.doi.org/}%
\providecommand \selectlanguage [0]{\@gobble}%
\providecommand \bibinfo  [0]{\@secondoftwo}%
\providecommand \bibfield  [0]{\@secondoftwo}%
\providecommand \translation [1]{[#1]}%
\providecommand \BibitemOpen [0]{}%
\providecommand \bibitemStop [0]{}%
\providecommand \bibitemNoStop [0]{.\EOS\space}%
\providecommand \EOS [0]{\spacefactor3000\relax}%
\providecommand \BibitemShut  [1]{\csname bibitem#1\endcsname}%
\let\auto@bib@innerbib\@empty
%</preamble>
\bibitem [{\citenamefont {Koschorreck}\ \emph {et~al.}(2013)\citenamefont
  {Koschorreck}, \citenamefont {Pertot}, \citenamefont {Vogt},\ and\
  \citenamefont {K{\"o}hl}}]{kohl2013}%
  \BibitemOpen
  \bibfield  {author} {\bibinfo {author} {\bibfnamefont {M.}~\bibnamefont
  {Koschorreck}}, \bibinfo {author} {\bibfnamefont {D.}~\bibnamefont {Pertot}},
  \bibinfo {author} {\bibfnamefont {E.}~\bibnamefont {Vogt}}, \ and\ \bibinfo
  {author} {\bibfnamefont {M.}~\bibnamefont {K{\"o}hl}},\ }\href@noop {}
  {\bibfield  {journal} {\bibinfo  {journal} {Nat.~Phys.}\ }\textbf {\bibinfo
  {volume} {9}},\ \bibinfo {pages} {405} (\bibinfo {year} {2013})}\BibitemShut
  {NoStop}%
\bibitem [{\citenamefont {Bardon}\ \emph {et~al.}(2014)\citenamefont {Bardon},
  \citenamefont {Beattie}, \citenamefont {Luciuk}, \citenamefont {Cairncross},
  \citenamefont {Fine}, \citenamefont {Cheng}, \citenamefont {Edge},
  \citenamefont {Taylor}, \citenamefont {Zhang}, \citenamefont {Trotzky},\ and\
  \citenamefont {Thywissen}}]{bardon2014transverse}%
  \BibitemOpen
  \bibfield  {author} {\bibinfo {author} {\bibfnamefont {A.~B.}\ \bibnamefont
  {Bardon}}, \bibinfo {author} {\bibfnamefont {S.}~\bibnamefont {Beattie}},
  \bibinfo {author} {\bibfnamefont {C.}~\bibnamefont {Luciuk}}, \bibinfo
  {author} {\bibfnamefont {W.}~\bibnamefont {Cairncross}}, \bibinfo {author}
  {\bibfnamefont {D.}~\bibnamefont {Fine}}, \bibinfo {author} {\bibfnamefont
  {N.~S.}\ \bibnamefont {Cheng}}, \bibinfo {author} {\bibfnamefont {G.~J.~A.}\
  \bibnamefont {Edge}}, \bibinfo {author} {\bibfnamefont {E.}~\bibnamefont
  {Taylor}}, \bibinfo {author} {\bibfnamefont {S.}~\bibnamefont {Zhang}},
  \bibinfo {author} {\bibfnamefont {S.}~\bibnamefont {Trotzky}}, \ and\
  \bibinfo {author} {\bibfnamefont {J.~H.}\ \bibnamefont {Thywissen}},\
  }\href@noop {} {\bibfield  {journal} {\bibinfo  {journal} {Science}\ }\textbf
  {\bibinfo {volume} {344}},\ \bibinfo {pages} {722} (\bibinfo {year}
  {2014})}\BibitemShut {NoStop}%
\bibitem [{\citenamefont {Trotzky}\ \emph {et~al.}(2015)\citenamefont
  {Trotzky}, \citenamefont {Beattie}, \citenamefont {Luciuk}, \citenamefont
  {Smale}, \citenamefont {Bardon}, \citenamefont {Enss}, \citenamefont
  {Taylor}, \citenamefont {Zhang},\ and\ \citenamefont
  {Thywissen}}]{trotzky2014observation}%
  \BibitemOpen
  \bibfield  {author} {\bibinfo {author} {\bibfnamefont {S.}~\bibnamefont
  {Trotzky}}, \bibinfo {author} {\bibfnamefont {S.}~\bibnamefont {Beattie}},
  \bibinfo {author} {\bibfnamefont {C.}~\bibnamefont {Luciuk}}, \bibinfo
  {author} {\bibfnamefont {S.}~\bibnamefont {Smale}}, \bibinfo {author}
  {\bibfnamefont {A.~B.}\ \bibnamefont {Bardon}}, \bibinfo {author}
  {\bibfnamefont {T.}~\bibnamefont {Enss}}, \bibinfo {author} {\bibfnamefont
  {E.}~\bibnamefont {Taylor}}, \bibinfo {author} {\bibfnamefont
  {S.}~\bibnamefont {Zhang}}, \ and\ \bibinfo {author} {\bibfnamefont {J.~H.}\
  \bibnamefont {Thywissen}},\ }\href {\doibase 10.1103/PhysRevLett.114.015301}
  {\bibfield  {journal} {\bibinfo  {journal} {Phys. Rev. Lett.}\ }\textbf
  {\bibinfo {volume} {114}},\ \bibinfo {pages} {015301} (\bibinfo {year}
  {2015})}\BibitemShut {NoStop}%
\bibitem [{\citenamefont {McGuirk}\ \emph {et~al.}(2002)\citenamefont
  {McGuirk}, \citenamefont {Lewandowski}, \citenamefont {Harber}, \citenamefont
  {Nikuni}, \citenamefont {Williams},\ and\ \citenamefont
  {Cornell}}]{mcguirk2002sw}%
  \BibitemOpen
  \bibfield  {author} {\bibinfo {author} {\bibfnamefont {J.~M.}\ \bibnamefont
  {McGuirk}}, \bibinfo {author} {\bibfnamefont {H.~J.}\ \bibnamefont
  {Lewandowski}}, \bibinfo {author} {\bibfnamefont {D.~M.}\ \bibnamefont
  {Harber}}, \bibinfo {author} {\bibfnamefont {T.}~\bibnamefont {Nikuni}},
  \bibinfo {author} {\bibfnamefont {J.~E.}\ \bibnamefont {Williams}}, \ and\
  \bibinfo {author} {\bibfnamefont {E.~A.}\ \bibnamefont {Cornell}},\
  }\href@noop {} {\bibfield  {journal} {\bibinfo  {journal} {Phys.~Rev.Lett.}\
  }\textbf {\bibinfo {volume} {89}},\ \bibinfo {pages} {090402} (\bibinfo
  {year} {2002})}\BibitemShut {NoStop}%
\bibitem [{\citenamefont {McGuirk}\ and\ \citenamefont
  {Zajiczek}(2011)}]{mcguirk2011localized}%
  \BibitemOpen
  \bibfield  {author} {\bibinfo {author} {\bibfnamefont {J.~M.}\ \bibnamefont
  {McGuirk}}\ and\ \bibinfo {author} {\bibfnamefont {L.~F.}\ \bibnamefont
  {Zajiczek}},\ }\href@noop {} {\bibfield  {journal} {\bibinfo  {journal}
  {Phys.~Rev.~A}\ }\textbf {\bibinfo {volume} {83}},\ \bibinfo {pages} {013625}
  (\bibinfo {year} {2011})}\BibitemShut {NoStop}%
\bibitem [{\citenamefont {Hild}\ \emph {et~al.}(2014)\citenamefont {Hild},
  \citenamefont {Fukuhara}, \citenamefont {Schau\ss}, \citenamefont {Zeiher},
  \citenamefont {Knap}, \citenamefont {Demler}, \citenamefont {Bloch},\ and\
  \citenamefont {Gross}}]{hild2014far}%
  \BibitemOpen
  \bibfield  {author} {\bibinfo {author} {\bibfnamefont {S.}~\bibnamefont
  {Hild}}, \bibinfo {author} {\bibfnamefont {T.}~\bibnamefont {Fukuhara}},
  \bibinfo {author} {\bibfnamefont {P.}~\bibnamefont {Schau\ss}}, \bibinfo
  {author} {\bibfnamefont {J.}~\bibnamefont {Zeiher}}, \bibinfo {author}
  {\bibfnamefont {M.}~\bibnamefont {Knap}}, \bibinfo {author} {\bibfnamefont
  {E.}~\bibnamefont {Demler}}, \bibinfo {author} {\bibfnamefont
  {I.}~\bibnamefont {Bloch}}, \ and\ \bibinfo {author} {\bibfnamefont
  {C.}~\bibnamefont {Gross}},\ }\href@noop {} {\bibfield  {journal} {\bibinfo
  {journal} {Phys.~Rev.~Lett.}\ }\textbf {\bibinfo {volume} {113}},\ \bibinfo
  {pages} {147205} (\bibinfo {year} {2014})}\BibitemShut {NoStop}%
\bibitem [{\citenamefont {Heinze}\ \emph {et~al.}(2013)\citenamefont {Heinze},
  \citenamefont {Krauser}, \citenamefont {Fl{\"a}schner}, \citenamefont
  {Sengstock}, \citenamefont {Becker}, \citenamefont {Ebling}, \citenamefont
  {Eckardt},\ and\ \citenamefont {Lewenstein}}]{heinze2013engineering}%
  \BibitemOpen
  \bibfield  {author} {\bibinfo {author} {\bibfnamefont {J.}~\bibnamefont
  {Heinze}}, \bibinfo {author} {\bibfnamefont {J.~S.}\ \bibnamefont {Krauser}},
  \bibinfo {author} {\bibfnamefont {N.}~\bibnamefont {Fl{\"a}schner}}, \bibinfo
  {author} {\bibfnamefont {K.}~\bibnamefont {Sengstock}}, \bibinfo {author}
  {\bibfnamefont {C.}~\bibnamefont {Becker}}, \bibinfo {author} {\bibfnamefont
  {U.}~\bibnamefont {Ebling}}, \bibinfo {author} {\bibfnamefont
  {A.}~\bibnamefont {Eckardt}}, \ and\ \bibinfo {author} {\bibfnamefont
  {M.}~\bibnamefont {Lewenstein}},\ }\href@noop {} {\bibfield  {journal}
  {\bibinfo  {journal} {Phys.~Rev.~Lett.}\ }\textbf {\bibinfo {volume} {110}},\
  \bibinfo {pages} {250402} (\bibinfo {year} {2013})}\BibitemShut {NoStop}%
\bibitem [{\citenamefont {Leggett}\ and\ \citenamefont
  {Rice}(1968)}]{Leggett1968}%
  \BibitemOpen
  \bibfield  {author} {\bibinfo {author} {\bibfnamefont {A.}~\bibnamefont
  {Leggett}}\ and\ \bibinfo {author} {\bibfnamefont {M.}~\bibnamefont {Rice}},\
  }\href {\doibase 10.1103/PhysRevLett.20.586} {\bibfield  {journal} {\bibinfo
  {journal} {Phys. Rev. Lett.}\ }\textbf {\bibinfo {volume} {20}},\ \bibinfo
  {pages} {586} (\bibinfo {year} {1968})}\BibitemShut {NoStop}%
\bibitem [{\citenamefont {Bashkin}(1981)}]{bashkin1981spin}%
  \BibitemOpen
  \bibfield  {author} {\bibinfo {author} {\bibfnamefont {E.~P.}\ \bibnamefont
  {Bashkin}},\ }\href@noop {} {\bibfield  {journal} {\bibinfo  {journal} {JETP
  Lett.}\ }\textbf {\bibinfo {volume} {33}},\ \bibinfo {pages} {8} (\bibinfo
  {year} {1981})}\BibitemShut {NoStop}%
\bibitem [{\citenamefont {Lhuillier}\ and\ \citenamefont
  {Lalo{\"e}}(1982{\natexlab{a}})}]{lhuillier1982transport}%
  \BibitemOpen
  \bibfield  {author} {\bibinfo {author} {\bibfnamefont {C.}~\bibnamefont
  {Lhuillier}}\ and\ \bibinfo {author} {\bibfnamefont {F.}~\bibnamefont
  {Lalo{\"e}}},\ }\href@noop {} {\bibfield  {journal} {\bibinfo  {journal}
  {J.~Physique}\ }\textbf {\bibinfo {volume} {43}},\ \bibinfo {pages} {197}
  (\bibinfo {year} {1982}{\natexlab{a}})}\BibitemShut {NoStop}%
\bibitem [{\citenamefont {Deutsch}\ \emph {et~al.}(2010)\citenamefont
  {Deutsch}, \citenamefont {Ramirez-Martinez}, \citenamefont {Lacro{\^u}te},
  \citenamefont {Reinhard}, \citenamefont {Schneider}, \citenamefont {Fuchs},
  \citenamefont {Pi{\'e}chon}, \citenamefont {Lalo{\"e}}, \citenamefont
  {Reichel},\ and\ \citenamefont {Rosenbusch}}]{reichel2010coherence}%
  \BibitemOpen
  \bibfield  {author} {\bibinfo {author} {\bibfnamefont {C.}~\bibnamefont
  {Deutsch}}, \bibinfo {author} {\bibfnamefont {F.}~\bibnamefont
  {Ramirez-Martinez}}, \bibinfo {author} {\bibfnamefont {C.}~\bibnamefont
  {Lacro{\^u}te}}, \bibinfo {author} {\bibfnamefont {F.}~\bibnamefont
  {Reinhard}}, \bibinfo {author} {\bibfnamefont {T.}~\bibnamefont {Schneider}},
  \bibinfo {author} {\bibfnamefont {J.}~\bibnamefont {Fuchs}}, \bibinfo
  {author} {\bibfnamefont {F.}~\bibnamefont {Pi{\'e}chon}}, \bibinfo {author}
  {\bibfnamefont {F.}~\bibnamefont {Lalo{\"e}}}, \bibinfo {author}
  {\bibfnamefont {J.}~\bibnamefont {Reichel}}, \ and\ \bibinfo {author}
  {\bibfnamefont {P.}~\bibnamefont {Rosenbusch}},\ }\href@noop {} {\bibfield
  {journal} {\bibinfo  {journal} {Phys.~Rev.~Lett.}\ }\textbf {\bibinfo
  {volume} {105}},\ \bibinfo {pages} {020401} (\bibinfo {year}
  {2010})}\BibitemShut {NoStop}%
\bibitem [{\citenamefont {Ragan}\ and\ \citenamefont
  {Schwarz}(1997)}]{ragan1997castaing}%
  \BibitemOpen
  \bibfield  {author} {\bibinfo {author} {\bibfnamefont {R.~J.}\ \bibnamefont
  {Ragan}}\ and\ \bibinfo {author} {\bibfnamefont {D.~M.}\ \bibnamefont
  {Schwarz}},\ }\href@noop {} {\bibfield  {journal} {\bibinfo  {journal}
  {J.~Low Temp.~Phys.}\ }\textbf {\bibinfo {volume} {109}},\ \bibinfo {pages}
  {775} (\bibinfo {year} {1997})}\BibitemShut {NoStop}%
\bibitem [{\citenamefont {Lhuillier}\ and\ \citenamefont
  {Lalo{\"e}}(1982{\natexlab{b}})}]{lhuillier1982transport2}%
  \BibitemOpen
  \bibfield  {author} {\bibinfo {author} {\bibfnamefont {C.}~\bibnamefont
  {Lhuillier}}\ and\ \bibinfo {author} {\bibfnamefont {F.}~\bibnamefont
  {Lalo{\"e}}},\ }\href@noop {} {\bibfield  {journal} {\bibinfo  {journal}
  {J.~Physique}\ }\textbf {\bibinfo {volume} {43}},\ \bibinfo {pages} {225}
  (\bibinfo {year} {1982}{\natexlab{b}})}\BibitemShut {NoStop}%
\bibitem [{\citenamefont {Mullin}\ and\ \citenamefont
  {Ragan}(2006)}]{mullin2006spin}%
  \BibitemOpen
  \bibfield  {author} {\bibinfo {author} {\bibfnamefont {W.~J.}\ \bibnamefont
  {Mullin}}\ and\ \bibinfo {author} {\bibfnamefont {R.~J.}\ \bibnamefont
  {Ragan}},\ }\href@noop {} {\bibfield  {journal} {\bibinfo  {journal} {Phys.
  Rev. A}\ }\textbf {\bibinfo {volume} {74}},\ \bibinfo {pages} {043607}
  (\bibinfo {year} {2006})}\BibitemShut {NoStop}%
\bibitem [{\citenamefont {Sommer}\ \emph {et~al.}(2011)\citenamefont {Sommer},
  \citenamefont {Ku}, \citenamefont {Roati},\ and\ \citenamefont
  {Zwierlein}}]{sommer2011universal}%
  \BibitemOpen
  \bibfield  {author} {\bibinfo {author} {\bibfnamefont {A.}~\bibnamefont
  {Sommer}}, \bibinfo {author} {\bibfnamefont {M.}~\bibnamefont {Ku}}, \bibinfo
  {author} {\bibfnamefont {G.}~\bibnamefont {Roati}}, \ and\ \bibinfo {author}
  {\bibfnamefont {M.~W.}\ \bibnamefont {Zwierlein}},\ }\href@noop {} {\bibfield
   {journal} {\bibinfo  {journal} {Nature}\ }\textbf {\bibinfo {volume}
  {472}},\ \bibinfo {pages} {201} (\bibinfo {year} {2011})}\BibitemShut
  {NoStop}%
\bibitem [{\citenamefont {Fuchs}\ \emph {et~al.}(2002)\citenamefont {Fuchs},
  \citenamefont {Gangardt},\ and\ \citenamefont
  {Lalo{\"e}}}]{fuchs2002internal}%
  \BibitemOpen
  \bibfield  {author} {\bibinfo {author} {\bibfnamefont {J.~N.}\ \bibnamefont
  {Fuchs}}, \bibinfo {author} {\bibfnamefont {D.~M.}\ \bibnamefont {Gangardt}},
  \ and\ \bibinfo {author} {\bibfnamefont {F.}~\bibnamefont {Lalo{\"e}}},\
  }\href {\doibase 10.1103/PhysRevLett.88.230404} {\bibfield  {journal}
  {\bibinfo  {journal} {Phys.~Rev.~Lett.}\ }\textbf {\bibinfo {volume} {88}},\
  \bibinfo {pages} {230404} (\bibinfo {year} {2002})}\BibitemShut {NoStop}%
\bibitem [{\citenamefont {Nikuni}\ \emph {et~al.}(2002)\citenamefont {Nikuni},
  \citenamefont {Williams},\ and\ \citenamefont {Clark}}]{nikuni2002linear}%
  \BibitemOpen
  \bibfield  {author} {\bibinfo {author} {\bibfnamefont {T.}~\bibnamefont
  {Nikuni}}, \bibinfo {author} {\bibfnamefont {J.~E.}\ \bibnamefont
  {Williams}}, \ and\ \bibinfo {author} {\bibfnamefont {C.~W.}\ \bibnamefont
  {Clark}},\ }\href {\doibase 10.1103/PhysRevA.66.043411} {\bibfield  {journal}
  {\bibinfo  {journal} {Phys. Rev. A}\ }\textbf {\bibinfo {volume} {66}},\
  \bibinfo {pages} {043411} (\bibinfo {year} {2002})}\BibitemShut {NoStop}%
\bibitem [{\citenamefont {Oktel}\ and\ \citenamefont
  {Levitov}(2002)}]{oktel2002internal}%
  \BibitemOpen
  \bibfield  {author} {\bibinfo {author} {\bibfnamefont {M.~{\"O}.}\
  \bibnamefont {Oktel}}\ and\ \bibinfo {author} {\bibfnamefont {L.~S.}\
  \bibnamefont {Levitov}},\ }\href@noop {} {\bibfield  {journal} {\bibinfo
  {journal} {Phys. Rev. Lett.}\ }\textbf {\bibinfo {volume} {88}},\ \bibinfo
  {pages} {230403} (\bibinfo {year} {2002})}\BibitemShut {NoStop}%
\bibitem [{\citenamefont {Williams}\ \emph {et~al.}(2002)\citenamefont
  {Williams}, \citenamefont {Nikuni},\ and\ \citenamefont
  {Clark}}]{williams2002longitudinal}%
  \BibitemOpen
  \bibfield  {author} {\bibinfo {author} {\bibfnamefont {J.~E.}\ \bibnamefont
  {Williams}}, \bibinfo {author} {\bibfnamefont {T.}~\bibnamefont {Nikuni}}, \
  and\ \bibinfo {author} {\bibfnamefont {C.~W.}\ \bibnamefont {Clark}},\ }\href
  {\doibase 10.1103/PhysRevLett.88.230405} {\bibfield  {journal} {\bibinfo
  {journal} {Phys. Rev. Lett.}\ }\textbf {\bibinfo {volume} {88}},\ \bibinfo
  {pages} {230405} (\bibinfo {year} {2002})}\BibitemShut {NoStop}%
\bibitem [{\citenamefont {McGuirk}\ and\ \citenamefont
  {Zajiczek}(2010)}]{mcguirk2010optical}%
  \BibitemOpen
  \bibfield  {author} {\bibinfo {author} {\bibfnamefont {J.~M.}\ \bibnamefont
  {McGuirk}}\ and\ \bibinfo {author} {\bibfnamefont {L.~F.}\ \bibnamefont
  {Zajiczek}},\ }\href@noop {} {\bibfield  {journal} {\bibinfo  {journal} {New
  J.~Phys.}\ }\textbf {\bibinfo {volume} {12}},\ \bibinfo {pages} {103020}
  (\bibinfo {year} {2010})}\BibitemShut {NoStop}%
\bibitem [{\citenamefont {Niroomand}(2013)}]{niroomand2013observation}%
  \BibitemOpen
  \bibfield  {author} {\bibinfo {author} {\bibfnamefont {D.}~\bibnamefont
  {Niroomand}},\ }\emph {\bibinfo {title} {Observation of the Castaing
  Instability in a Trapped Ultracold Bose Gas}},\ \href@noop {} {Master's
  thesis},\ \bibinfo  {school} {Simon Fraser University} (\bibinfo {year}
  {2013})\BibitemShut {NoStop}%
\bibitem [{\citenamefont {Nunes}\ \emph {et~al.}(1992)\citenamefont {Nunes},
  \citenamefont {Jin}, \citenamefont {Hawthorne}, \citenamefont {Putnam},\ and\
  \citenamefont {Lee}}]{nunes1992spin}%
  \BibitemOpen
  \bibfield  {author} {\bibinfo {author} {\bibfnamefont {G.}~\bibnamefont
  {Nunes}}, \bibinfo {author} {\bibfnamefont {C.}~\bibnamefont {Jin}}, \bibinfo
  {author} {\bibfnamefont {D.~L.}\ \bibnamefont {Hawthorne}}, \bibinfo {author}
  {\bibfnamefont {A.~M.}\ \bibnamefont {Putnam}}, \ and\ \bibinfo {author}
  {\bibfnamefont {D.~M.}\ \bibnamefont {Lee}},\ }\href {\doibase
  10.1103/PhysRevB.46.9082} {\bibfield  {journal} {\bibinfo  {journal} {Phys.
  Rev. B}\ }\textbf {\bibinfo {volume} {46}},\ \bibinfo {pages} {9082}
  (\bibinfo {year} {1992})}\BibitemShut {NoStop}%
\bibitem [{\citenamefont {Akimoto}\ \emph {et~al.}(1991)\citenamefont
  {Akimoto}, \citenamefont {Ishikawa}, \citenamefont {Oh}, \citenamefont
  {Nakagawa}, \citenamefont {Hata},\ and\ \citenamefont
  {Kodama}}]{akimoto1991nonlinear}%
  \BibitemOpen
  \bibfield  {author} {\bibinfo {author} {\bibfnamefont {H.}~\bibnamefont
  {Akimoto}}, \bibinfo {author} {\bibfnamefont {O.}~\bibnamefont {Ishikawa}},
  \bibinfo {author} {\bibfnamefont {G.-H.}\ \bibnamefont {Oh}}, \bibinfo
  {author} {\bibfnamefont {M.}~\bibnamefont {Nakagawa}}, \bibinfo {author}
  {\bibfnamefont {T.}~\bibnamefont {Hata}}, \ and\ \bibinfo {author}
  {\bibfnamefont {T.}~\bibnamefont {Kodama}},\ }\href@noop {} {\bibfield
  {journal} {\bibinfo  {journal} {J.~Low Temp.~Phys.}\ }\textbf {\bibinfo
  {volume} {82}},\ \bibinfo {pages} {295} (\bibinfo {year} {1991})}\BibitemShut
  {NoStop}%
\bibitem [{\citenamefont {K{\"o}nig}\ \emph {et~al.}(1995)\citenamefont
  {K{\"o}nig}, \citenamefont {Ager}, \citenamefont {Bowley}, \citenamefont
  {Owers-Bradley},\ and\ \citenamefont {Meyerovich}}]{konig1995spin}%
  \BibitemOpen
  \bibfield  {author} {\bibinfo {author} {\bibfnamefont {R.}~\bibnamefont
  {K{\"o}nig}}, \bibinfo {author} {\bibfnamefont {J.~H.}\ \bibnamefont {Ager}},
  \bibinfo {author} {\bibfnamefont {R.~M.}\ \bibnamefont {Bowley}}, \bibinfo
  {author} {\bibfnamefont {J.~R.}\ \bibnamefont {Owers-Bradley}}, \ and\
  \bibinfo {author} {\bibfnamefont {A.~E.}\ \bibnamefont {Meyerovich}},\
  }\href@noop {} {\bibfield  {journal} {\bibinfo  {journal} {J.~Low
  Temp.~Phys.}\ }\textbf {\bibinfo {volume} {101}},\ \bibinfo {pages} {833}
  (\bibinfo {year} {1995})}\BibitemShut {NoStop}%
\bibitem [{\citenamefont {Ragan}\ and\ \citenamefont
  {Baggett}(2004)}]{ragan2004leggett}%
  \BibitemOpen
  \bibfield  {author} {\bibinfo {author} {\bibfnamefont {R.}~\bibnamefont
  {Ragan}}\ and\ \bibinfo {author} {\bibfnamefont {J.}~\bibnamefont
  {Baggett}},\ }\href@noop {} {\bibfield  {journal} {\bibinfo  {journal}
  {J.~Low Temp.~Phys.}\ }\textbf {\bibinfo {volume} {134}},\ \bibinfo {pages}
  {369} (\bibinfo {year} {2004})}\BibitemShut {NoStop}%
\bibitem [{\citenamefont {Kuklov}\ and\ \citenamefont
  {Meyerovich}(2002)}]{kuklov2002precessing}%
  \BibitemOpen
  \bibfield  {author} {\bibinfo {author} {\bibfnamefont {A.}~\bibnamefont
  {Kuklov}}\ and\ \bibinfo {author} {\bibfnamefont {A.~E.}\ \bibnamefont
  {Meyerovich}},\ }\href {\doibase 10.1103/PhysRevA.66.023607} {\bibfield
  {journal} {\bibinfo  {journal} {Phys. Rev. A}\ }\textbf {\bibinfo {volume}
  {66}},\ \bibinfo {pages} {023607} (\bibinfo {year} {2002})}\BibitemShut
  {NoStop}%
\bibitem [{\citenamefont {Fuchs}\ \emph {et~al.}(2003)\citenamefont {Fuchs},
  \citenamefont {Pr\'{e}vot\'{e}},\ and\ \citenamefont
  {Gangardt}}]{fuchs2003castaing}%
  \BibitemOpen
  \bibfield  {author} {\bibinfo {author} {\bibfnamefont {J.~N.}\ \bibnamefont
  {Fuchs}}, \bibinfo {author} {\bibfnamefont {O.}~\bibnamefont
  {Pr\'{e}vot\'{e}}}, \ and\ \bibinfo {author} {\bibfnamefont {D.~M.}\
  \bibnamefont {Gangardt}},\ }\href@noop {} {\bibfield  {journal} {\bibinfo
  {journal} {Eur.~Phys.~J.~D}\ }\textbf {\bibinfo {volume} {25}},\ \bibinfo
  {pages} {167} (\bibinfo {year} {2003})}\BibitemShut {NoStop}%
\bibitem [{\citenamefont {Meyerovich}(1989)}]{meyerovich1989quantum}%
  \BibitemOpen
  \bibfield  {author} {\bibinfo {author} {\bibfnamefont {A.~E.}\ \bibnamefont
  {Meyerovich}},\ }\href {\doibase 10.1103/PhysRevB.39.9318} {\bibfield
  {journal} {\bibinfo  {journal} {Phys. Rev. B}\ }\textbf {\bibinfo {volume}
  {39}},\ \bibinfo {pages} {9318} (\bibinfo {year} {1989})}\BibitemShut
  {NoStop}%
\end{thebibliography}%

\end{document}